# Angular behavior of the absorption limit in thin film silicon solar cells

Ali Naqavi[a,b], Franz-Josef Haug[a], Karin Söderström[a], Corsin Battaglia[a], Vincent Paeder[b], Toralf Scharf[b], Hans Peter Herzig[b], Christophe Ballif[a]

[a]Photovoltaics and Thin Film Electronics Laboratory, Institute of Microengineering (IMT), École Polytechnique Fédérale de Lausanne (EPFL), Rue A.-L. Breguet 2, 2000 Neuchâtel, Switzerland
[b]Optics & Photonics Technology Laboratory, École Polytechnique Fédérale de Lausanne (EPFL), Rue A.-L. Breguet 2, 2000 Neuchâtel, Switzerland

## Abstract
We investigate the angular behavior of the upper bound of absorption provided by the guided modes in thin film solar cells. We show that the $4n^2$ limit can be potentially exceeded in a wide angular and wavelength range using two-dimensional periodic thin film structures. Two models are used to estimate the absorption enhancement; in the first one, we apply the periodicity condition along the thickness of the thin film structure but in the second one, we consider imperfect confinement of the wave to the device. To extract the guided modes, we use an automatized procedure which is established in this work. Through examples, we show that from the optical point of view, thin film structures have a high potential to be improved by changing their shape. Also, we discuss the nature of different optical resonances which can be potentially used to enhance light trapping in the solar cell. We investigate the two different polarization directions for one-dimensional gratings and we show that the transverse magnetic polarization can provide higher values of absorption enhancement. We also propose a way to reduce the angular dependence of the solar cell efficiency by the appropriate choice of periodic pattern. Finally, to get more practical values for the absorption enhancement, we consider the effect of parasitic loss which can significantly reduce the enhancement factor.

## *1. Introduction*

Thin film silicon technology has attracted a lot of attention in the previous decades as a major candidate for solar energy harvesting. Despite the abundance of silicon, reduced device thicknesses are preferable due to higher production throughput and better device stability. Feasible devices are typically thinner than the optical absorption length almost regardless of the type of silicon in a significant part of the solar spectrum above the bandgap [1-4]. Therefore, methods for enhancing light absorption inside the device, so called light trapping strategies, become necessary to obtain thin film silicon solar cells with high stabilized efficiencies. Light trapping can be understood as the way to couple light from outside the cell to the modes inside the structure and to absorb it as much as possible before it escapes outside.

From an experimental point of view, the most widely used light trapping concept makes use of randomly textured interfaces [5-7]. Theoretical considerations suggest that the maximal amount of absorption enhancement compared to the single pass absorption is limited by the refractive index of the absorber material, $n$, through the simple relation $4n^2$ [8]. This limit, proposed by Yablonovitch and Cody, is based on three major assumptions: first, light absorption is weak in the device; second, the interfaces are ideal Lambertian scatterers which couple uniformly to the continuum of electromagnetic modes within the slab regardless of the angle of incidence of the incoming radiation and third, the absorber layer is thick so that geometrical optics approximations can be used. In real thin film devices, the mentioned assumptions do not apply. For example, the interfaces do not scatter light into different directions uniformly. This introduces a degree of freedom to surpass the $4n^2$ limit; restricting the angle of incidence and escape to angles below $\theta$, can increase the enhancement factor up to $4n^2/\sin^2(\theta)$ [8, 9]. Following this intuition, it has been proposed that it is possible to surpass the $4n^2$ limit in a desired part of the spectrum using grating patterns [9, 10]. Gratings can couple the incident light more effectively to certain directions (wave-vectors) which enhance absorption more significantly over the wavelength range of interest. Recently, it was shown theoretically through a statistical approach that it may be possible to increase the enhancement factor up to $\frac{8}{\sqrt{3}}\pi n^2$ over a limited wavelength range [9]. However, the supporting calculations are based on the assumption of thick absorber layers. In the case of real thin film devices, thick layers do not apply and the method should be modified [11].

Light absorption in solar cells is a function of incident angle, regardless of the type of texturing. Unfortunately, most of solar cell literature deals with the case of normal incidence of light. Recently, Yu *et al.* proposed a model to determine the light trapping limit for arbitrary angles of incidence [9]. In their model, Yu *et al.* assumed an absorber film that is thick enough to envelop discrete modes by a continuum model. However, recent work on thin film devices with periodically textured interfaces showed that discrete waveguide modes can be distinguished, and that the excitation of these modes varies significantly with the angle of incidence [12-14]. In this paper, we investigate the upper limit of absorption enhancement in a thin film solar cell based on gratings, taking into account specifically the angular dependence of this limit. The cell thickness is considered only a few hundreds of nanometers to simulate the case of a hydrogenated amorphous silicon (a-Si:H) solar cell and, hence, to enable acceptable stability against light induced degradation.

Angular behavior of light absorption in thin film devices mainly depends on interference and resonant excitation of guided modes. Specifically, resonances are discussed in this work by taking into account the following properties; first, the field of a guided mode supported by a thin dielectric film can extend over a considerable distance out of the actual guiding medium. This effect is also present in thick films but in this case, the decay length into air is negligible compared to the film thickness. Second, the interface texture serves to couple energy from the incident light into a given mode. In this contribution, we treat the interface texture in terms of periodic perturbations which permit the application of band folding to the dispersion relation of the corresponding flat interface model [15].

One main assumption behind the calculation of the upper limit of absorption enhancement is that after coupling to any specific mode, the field energy is totally absorbed and converted into photocurrent. In practice, there are always such deficiencies as parasitic loss and the spread of energy over space [16] which deteriorate the enhancement. We calculate the impact of both of the mentioned phenomena by using the solution of a multilayer stack with flat interfaces which is taken as the reference. Since solution of Maxwell equations for the reference structure is relatively easy, the overlap integrals concerning the spatial distribution of energy and absorption can be calculated efficiently. Clearly, if the structure is very different from the planar reference, the mentioned approximation becomes less accurate.

This paper is arranged as follows. In section 2, the slab waveguide model is explained and the different nature of its various types of resonances is discussed. Radiation modes and guided modes can provide resonances which increase light absorption in the solar cell. Their angular dependence is also discussed in section 2. In section 3, the effect of periodicity on the guided modes of a structure is included assuming also periodicity along the cell thickness for simplicity. Section 3 deals only with slab waveguides to make the analysis as easy as possible and it describes the impact of device thickness on the enhancement factor. In Section 4, we include both periodicity and the open nature of the guide. For this, we need a procedure to extract the modes of the planar structure and to count the corresponding resonances. This procedure is explained in Appendix A. In Appendix B we describe how to calculate the enhancement factor over a wide wavelength range while considering the open waveguide condition. Section 5 explains the upper bound of absorption enhancement for a complete solar cell stack and also it includes a general discussion on the effect of incident light polarization, the impact of resonances on the angular behavior of the absorption and the extension of the analysis to the case of 2D gratings. To find more practical values for the absorption enhancement, the loss due to parasitic absorption should be deduced from the absorbed power. The latter modification is applied in section 6. Though this correction does not introduce a fundamental limitation on the absorption enhancement, it provides a more applied approximation to be consistent with the real world results. Finally, in section 7, we highlight some practical considerations and conclusions. For example, we obtain an approximation of the maximal photocurrent enhancement provided by a 1D grating pattern.

## *2. Angular dependence of resonances: slab waveguide model*

The mechanism of absorption enhancement can be understood by a perturbation approach where the solar cell stack is regarded as a planar waveguide. In such a configuration, interface texture can be considered as a perturbation that establishes the coupling from external radiation to waveguide modes, and the assumption of periodic textures is particularly convenient for the description [15, 17]. The approach remains valid for more complicated interface textures via spatial Fourier analysis.

In Figure 1, a dielectric slab waveguide and its dispersion diagram are shown. The dispersion diagram indicates the values of the parallel component of the incident wave vector k and photon energy E which result in constructive interference inside the slab. In other words, it shows the $(k_\parallel, E)$ pairs which correspond to the guided modes of the slab. The slab thickness is 200 nm and its refractive index is n=4. The dispersion diagram is plotted for the case where the electric field vector is in *y* direction and does not have a *z* component (TE). Similar qualitative behavior is obtained for transverse magnetic fields (TM). One can divide the dispersion diagram into three regions; the region above of air light line which is a continuum of *radiation modes*, the part between the light lines of air (*n*=1) and the dielectric (*n*=4) in which the guided modes of the structure exist and the region below the light line of the dielectric which does not support any modes. Introduction of a metal cladding can produce a surface plasmon mode whose dispersion relation goes underneath the dielectric light line. For the case of metallic solar cell back reflectors, the surface plasmon mode is normally suppressed by a dielectric buffer layer [15].

Resonances due to radiation modes are Fabry-Pérot like interferences. They originate from the constructive interference of the waves which are transmitted through or reflected from the top and the bottom interfaces of the slab. As the refractive index of the slab is larger than that of air, the wave vector of the radiation modes will be refracted towards the surface normal while entering the dielectric according to Snell's law of refraction. If the incident angle is increased to grazing incidence, the refraction angle will represent the critical angle of the dielectric; in the graph of Figure 1, it will correspond to a point on the light line of air. Refraction angles larger than the critical angle cannot be associated with a physical incident angle. If such angles satisfy the resonance condition inside the dielectric, they may represent guided modes.

Guided modes show a different behavior in terms of energy coupling than the radiation modes; while radiation modes are not bound to the optical device, an ideal guided wave is completely trapped inside its corresponding geometry, i.e. it does not have any exchange of energy with the outside environment. It can neither be excited from outside, nor can it lose energy by coupling energy to the waves radiating into air. Any energy imparted to a guided mode, can stay inside the waveguide for a long time - ideally forever. If an absorber is placed inside, or if the guiding medium itself is absorbing, all of the input energy will be absorbed in the waveguide.

In solar cells, coupling to guided modes is achieved by the interface texture. The amounts of in-coupling and out-coupling to a resonating mode are not independent; light that can be coupled from a radiating field into a resonance can also be coupled from the resonance to the outside medium. In a solar cell, in-coupling of light is desired to be as strong as possible and out-coupling is desired to be zero. Hence, the light trapping scheme should benefit from the input waves and absorb them as much as possible before they escape outside. A basic assumption in this contribution is that the dielectrics are considered weakly absorbing as in [9]. Therefore, we assume that the main limitation in enhancing absorption in the solar cell is the weak absorption of the material and not coupling light from outside to the guided modes [9]. Thus, we approximate the upper bound of the absorption enhancement. In practice, however, the actual amount of coupling to a given guided mode should be considered [18].

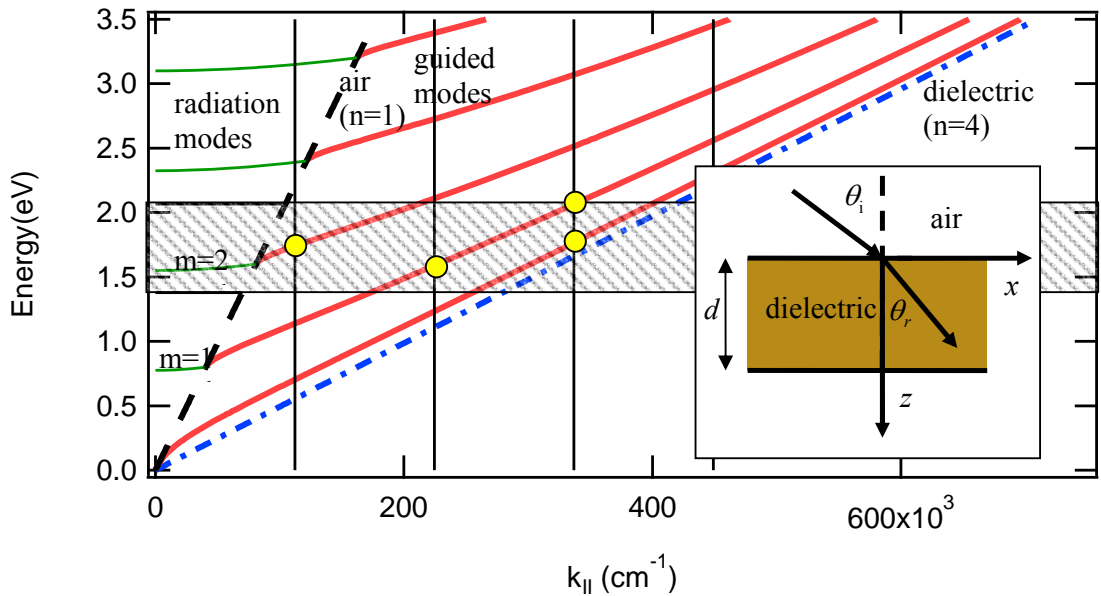

**Figure 1-** (Color online) A typical slab waveguide (inset) and its dispersion diagram. The electric field vector is supposed to be parallel to the slab interfaces. The thick dotted-dashed and dashed lines show the light line of the dielectric and the light line of air respectively. The red lines represent the guided modes. The thin green lines show the interference maxima. The continuum of radiation modes falls in the region between the energy axis and the light line of air. The two horizontal black solid lines and the shaded area between them show the energy range equivalent to the wavelength range of interest. The values of *m* beside the lines of guided modes show the mode number associated to Eq. (1). The vertical lines correspond to the first, second and third order diffraction orders. Small circles mark excitation of guided modes by the grating pattern.

The guided modes show more dramatic angular changes than the interferences. The latter observation can be understood by the ray optics model; guided modes are associated with beams that illuminate the slab-air interface from inside the slab at angles beyond the critical angle; since the effective light path length increases more at larger angles, guided modes show more angular variations. To be more specific, we can derive the resonant wavelengths $\lambda$ of a parallel plate waveguide as an approximation [19] - Figure 1.

$$\lambda = \frac{2nd \cos \theta_r}{m}. \tag{1}$$

where *n* is the refractive index of the material in the slab, *m* stands for the order of resonance and $\theta_r$ represents the refraction angle. In the wave optics regime and especially if one is going to distinguish polarization, *d* should be modified to include penetration of the wave into air. In our analysis we focus on a-Si:H cells. Light trapping is thus mainly required in regard with red light since amorphous silicon absorbs red light weakly. So, it is desired to have as many resonances as possible in the red part of the spectrum shown by the shaded area in Figure 1. Eq. (1) clearly shows that at smaller angles, the resonant wavelength takes higher values. Due to Snell's law, $\sin \theta_i = n \sin \theta_r$, and Eq. (1), a given radiation mode of order *m* that is associated to the red light for small $\theta_i$ (close to perpendicular) is thus shifted towards blue if $\theta_i$ is increased. Similarly, a resonance due to a guided mode will shift towards the short wavelength region by increasing $\theta_r$. Because of these angular changes, some of the resonances might escape from the wavelength range of interest at large angles. By using Snell's law and the identity $\cos^2 \theta_r + \sin^2 \theta_r = 1$, we can rewrite Eq. (1) as

$$k_\parallel = k_0 \sin \theta_i = nk_0 \sqrt{1 - \left(\frac{m\lambda}{2nd}\right)^2} \tag{2}$$

Radiation modes are identified by $k_\parallel$ values smaller than $k_0$ while the guided modes are associated with $k_\parallel > k_0$. By changing the wavelength in Eq. (2), resonances due to the radiation mode excitation can be determined. In Figure 1, the curves related to m =0,1,2,3,4 are shown. The wavelength range of interest is from 600 nm (2.07 eV) to 900 nm (1.38 eV). Figure 1 shows that only m = 0,1,2 can satisfy Eq. (2) within the considered wavelength range. At near normal incidence, just the m= 2 interference is excited but as $k_\parallel$ is increased, the m=0,1 modes enter the desired wavelength range and the m=2 mode leaves it.

### 3. Modeling periodicity

Using periodic structures can be beneficial to enhance light trapping. As observed in both theory [13] and experiment [12], periodic patterns e.g. gratings provide a means to couple light form outside the structure to its guided modes. Periodicity results in folding of bands at the edges of the Brillouin zones in the dispersion diagram which is equivalent to coupling to various diffraction orders. As an example, assume that the film with the modal dispersion plotted in Figure 1 is slightly perturbed by a 1D grating pattern with period P=560 nm. We use this period to be consistent with our previous work [13]. The vertical lines correspond to the diffraction orders for normal incidence and the small circles which occur at the crossings of the guided mode curves and the diffraction lines show the excitation of guided modes in the wavelength range of interest. An illuminating plane wave can potentially satisfy the phase matching condition of the same mode in more than one wavelength. For example, coupling of the normally incident plane wave to the third order diffraction can excite the two lowest order modes at two different wavelengths. This is shown in Figure 1 by two circles on the same vertical line which is attributed to the third order diffraction. In the mentioned example, four resonances can be excited by grating-assisted coupling. In this way, periodicity increases the number of resonances in the spectral domain. As discussed in the previous section, by changing the incident angle, the resonant wavelength of some of the modes might shift outside the wavelength range due to the change of Bragg condition, i.e. shift of the vertical lines in Figure 1. The periodic pattern can satisfy the phase matching condition of other guided modes corresponding to other diffraction orders instead; this is equivalent to changing the Brillouin zone in the dispersion diagram. Hence, the number of resonances does not vary as significantly as the case with flat interfaces. So, the periodic structure serves to reduce the dependence of photocurrent generation on the incident angle.

Periodicity can be modeled by considering the resonance conditions in k-space as applied by Yu *et al.* [9]. This approach is based on the assumption of periodicity not only in xy plane but also along the z – direction since they model relatively thick solar cells. Therefore, the resonances that a plane wave can excite, satisfy two main criteria; first, the phase matching condition in the xy plane and second, transverse resonance in z direction, i.e. $k_z = 2\pi m/d$. We will refer to the latter model as the "idealized model" from now on. The "idealized model" is equivalent to applying "closed waveguide" boundary condition along z-direction. It provides acceptable results for solar cells with thick layers since in that case, the field penetration length into air is negligible compared to the guide thickness, *d*. Thin films, however, require a modified model which can consider also the effect of field confinement in the transverse direction. Yu *et al.* have previously discussed very briefly the thin film solar cells by introducing a correction factor [20] which describes the field overlap with the device but they did not explain in detail how to find that factor. We perform the latter correction by folding the dispersion diagram of the *open* waveguide. In this fashion, we modify the results of [9] for thin films by taking into account the important property of dealing with an "open" guided wave structure for the thin film geometry. We refer to the latter model as the "real model". In the "real model", a resonance must satisfy phase matching in xy plane similar to the "idealized model"; however, the transverse resonance condition should be modified. We apply this modification by using the dispersion diagram of the multilayer structure with flat interfaces.

The absorption enhancement factor can be introduced as the figure of merit to evaluate the amount of absorption enhancement, so called "light path length enhancement". It is defined for each structure as the ratio of the absorption to the single-pass absorption and the structure is supposed to be weakly absorbing [9]. The maximum enhancement factor can be obtained through the following relation [9].

$$F = \frac{c}{nd\Delta f}\frac{\Delta M}{N}. \tag{3}$$

in which *n*, *d* and *c* are the refractive index, thickness of the guide and the free space speed of light respectively. Δ*M* stands for the number of allowed guided mode resonances in the frequency interval Δ*f* [21] and *N* refers to the number of radiating reflection orders of the grating, so called channels. The number of resonances and the number of channels depend on the wavelength and the incident angle. Obviously, they also depend on the band diagram of the structure. Procedures for the evaluation of *F* were outlined for the idealized model and for the realistic model in Ref. [9] and [11] respectively. It was previously shown that in thick structures and over a wide spectral range, *F* can reach up to π*n* if 1D periodicity is applied [9] and up to 4$n^2$ in the general case- which also includes the 2D gratings [8, 9]. In a narrow wavelength and angular range, the maximal enhancement values for the thick structures are $2\pi n$ for 1D periodic structures and $4\pi n^2$ and $8\pi n^2/\sqrt{3}$ for 2D periodic structures with square and triangular periodicities. According to Eq. (3), the enhancement factor increases with the decrease in the number of channels. This can justify the use of subwavelength gratings since they support one order of reflection, hence, $N=1$ in Eq. (3). However, it should be mentioned that the period of the grating should not be much smaller than the operation wavelengths since in this case the electromagnetic field will not feel the change of geometry and from an experimental point of view, fabrication might be challenging. Note that in a real device, the mode is not fully confined to the active layer. The absorption in the active layer is thus enhanced less than the values predicted by Eq. (3). Spread of the mode profile over other layers e.g. the doped layers mainly results in parasitic absorption and therefore does not lead to improvement of photocurrent. We will come back to this point later in section 6.

Before discussing examples, it is beneficial to define the two different polarizations regarding a 1D grating. The P and the S polarizations refer to the cases where the electric and the magnetic field vectors are parallel to the grating grooves respectively. All examples correspond to the 1D grating case unless otherwise specified. Periodicity is applied in *x*-direction for 1D gratings and in *xy* plane for 2D gratings.

As the first example, we consider an amorphous silicon slab. The period is taken equal to 560 *nm* as in the experimental case in our previous works [12, 13] and the wavelength range extends from 600 to 900 *nm*. For any fixed geometry, the enhancement factor depends on the incident angle and the wavelength. Therefore, we can take the average value of the enhancement factor over the wavelength range to have an effective enhancement factor for each incident angle. This averaging can be best performed by using an appropriate weighting function e.g. the solar spectrum. However, the previous approximations of the upper bound of the absorption enhancement do not distinguish different wavelengths [8, 9]. We thus perform a simple averaging over wavelength to ease comparing the obtained results with the existing literature. Later in section 7, we introduce the impact of the solar spectrum on the results while calculating the photocurrent enhancement. Figure 2 shows the results of our calculation of enhancement factor for a 1D grating based textured slab on "the idealized model" for the case where refractive index of silicon is taken $n = 4$ (with the extinction coefficient of k=0) and the case in which we use the refractive index dispersion of a-Si:H. For simplicity, on axis incidence is considered only i.e. the incidence occurs in a plane normal to the grating grooves. At the weak absorption limit, both single pass absorption and the maximal absorption depend linearly on the absorption coefficient α. The enhancement factor in Eq. (3) can thus be obtained independent of the absorption coefficient α and the extinction coefficient k. The results are shown for thicknesses of 200 and 355 nm. The thickness of 355 nm is considered since it is equal to the thickness of the complete solar cell stack which is to be discussed later in an example. For the sake of simplicity, we use the same values of 1D and 2D limit throughout the paper which are calculated for $n = 4$; the 1D and 2D wide band limits are $\pi n$=12.57 and $4n^2$=64 respectively. The narrow band 1D limit of $2\pi n$=25.13 is also shown in the figure. It is observed in Figure 2 that the enhancement factor is greater for thinner slabs over the whole angular range. This can be attributed to the stronger single pass absorption in thick devices. The enhancement factor is maximal at around normal incidence but it does not reach the 2D wideband limit. As evidenced by the two curves corresponding to the 355 nm thick device, slight perturbations of the refractive index changes the enhancement factor more dramatically if the structure is thin. This can be explained by noting that if the device is thick, the single pass absorption will be strong and changing the refractive index will not change the total single pass absorption very much. This is contrary to the case of thin films where single pass absorption is so small that a slight variation in the refractive index results in dramatic changes of absorption. The enhancement factor *F* approaches the $\pi n$ line at large incident angles if the cell is thin; however, it surpasses this limit under near normal illumination even if the film is thick.

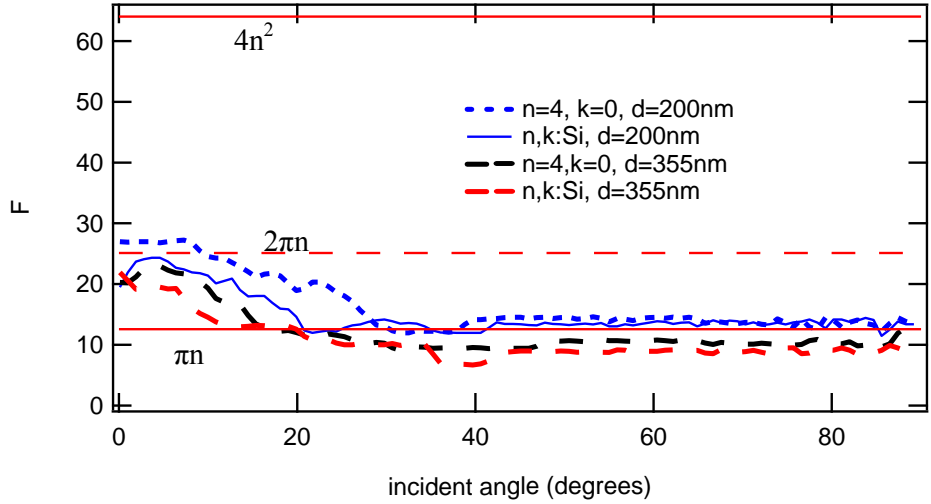

**Figure 2-** (Color online) Enhancement factor of a dielectric slab averaged over the wavelengths from 600 to 900 nm as a function of incident angle for 1D grating with period of 560 nm. The horizontal lines represent the limit for 1D and 2D gratings. The other four curves represent the enhancement factor which is obtained by the "idealized" model for the 200 nm thick slab with *n*=4 (dotted blue), the 200 nm thick a-Si:H slab (solid blue), the 355 nm thick slab with *n*=4 (dashed black) and the 355 nm thick a-Si:H slab (dashed red). For simplicity, on axis incidence is considered only.

## *4. The open nature of the thin film*

As previously mentioned, the assumption of periodicity in *z* direction is not valid in the case of open waveguides since field penetration into air can be comparable to the device thickness. The latter point leads to the increased effective thickness of the waveguide which based on Eq. (1), leads to compression of the dispersion diagram towards lower energies. This, in turn, means a higher density of guided modes and therefore, higher potential for absorption which will be verified in the next example.

In the first example, the idealized model was used to determine the absorption enhancement factor. In the second example, we find the enhancement factor of a 355 nm thick amorphous silicon slab using both "idealized" and "real" models. While applying the "real" model, to take into account that the structure under investigation is an "open" waveguide and not a "closed" one, we use the dispersion diagram of the structure with flat interfaces as an approximation instead of considering the transverse resonance condition $k_z = m\frac{2\pi}{d}$. We have recently utilized this "real model" to investigate thin film solar cells under normal incident light [11]. To consider other angles than normal, we need to modify both number of channels *N* and number of resonances $\Delta M$. Counting the number of channels is the same for both "idealized" and "real" models but counting $\Delta M$ needs more consideration which we explain in Appendix A. The procedure to calculate F in the "real model" is explained in Appendix B. Figure 3 shows three curves corresponding to the "real model": S- and P polarization and also the unpolarized result which is found by averaging the enhancement factor in the two polarizations. In the "idealized model", change of polarization does not vary the result. The values of the enhancement factor obtained through the "real model" are larger than the ones of the "idealized" one, regardless of polarization. This can be understood by the increased effective thickness of the dielectric waveguide which is considered in the "real model" or equivalently, lower effective guide index compared to its physical refractive index [22]. Furthermore, for incident angles smaller than 25°, the S polarized field leads to greater enhancement factor values compared to the P polarized field. We will see later the same phenomenon in a complete solar cell stack.

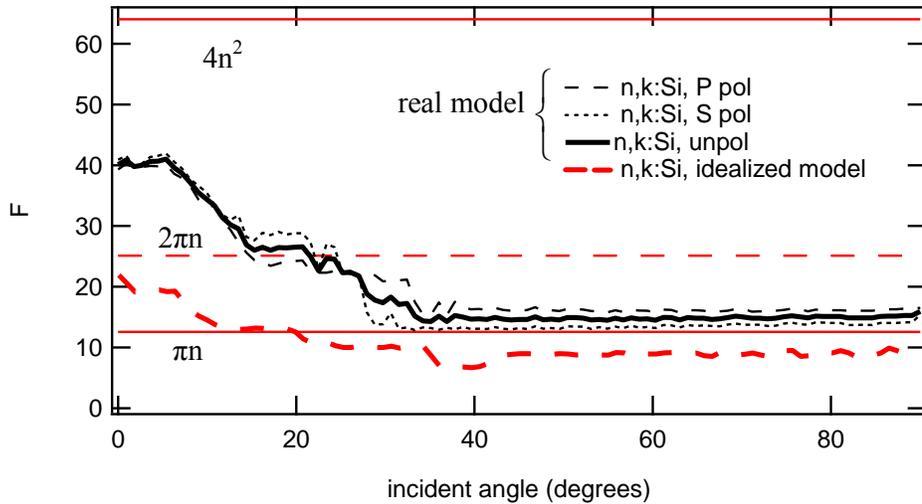

**Figure 3-** (Color online) Enhancement factor of a dielectric slab averaged over the wavelengths from 600 to 900 nm obtained by the "real" and the "idealized" models as a function of incident angle for 1D gratings. The horizontal lines represent the limit for 1D and 2D gratings. The other curves represent the enhancement factor for the 355 nm thick a-Si:H slab obtained with the idealized model (bold dashed red) and the real model in P polarization (dashed black), S polarization (dotted black) and the unpolarized case (bold black). For simplicity, on axis incidence is considered only.

## *5. Complete solar cell stack*

As the next example, the maximal enhancement factor is determined based on the "real model" for a complete solar cell stack on a 1D grating substrate including 70 nm of ITO front contact electrode, 15 nm of p-doped silicon (p-Si), 180 nm of intrinsic amorphous silicon (i-Si), 20 nm of n-doped silicon (n-Si), and 70 nm of ZnO respectively from top to bottom. The existence of different media in the same waveguide introduces an ambiguity in the application of Eq. (3). In the denominator of (3), the refractive index of the guide should be replaced with an effective refractive index $n_{eff} = k_{\parallel} / k_0$ [22]. The results are plotted in Figure 4 . The back reflector of the cell is considered to be silver. Again the P polarized, the S polarized and the unpolarized results are presented. The enhancement factor of the complete cell shows a similar trend as the enhancement factor of the a-Si:H slab under unpolarized illumination found with the "real model" but it is smaller than that in the whole spectral range. Since the counting model is identical in both cases, their difference can be related to the change of refractive indices; the higher refractive index of the silicon compared to the other layers leads to higher density of modes if the whole structure is made of silicon.

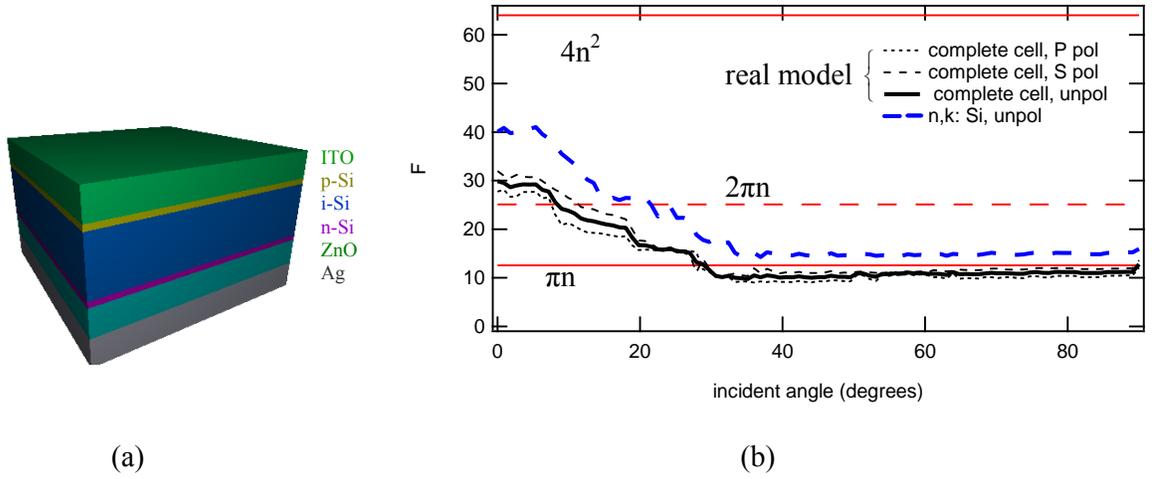

(a) (b)

**Figure 4-** (Color online) (a): layers of the full functional solar cell stack including doped and contact layers. (b): Comparison of the average enhancement factor of the solar cell structure and a slab of a-Si:H as a function of incident angle for 1D gratings. The horizontal lines represent the limit for 1D and 2D gratings. All of the other curves are obtained with the real model and they represent the enhancement factor obtained for the 355 nm thick a-Si:H slab under unpolarized illumination (bold dashed blue) and for the complete solar cell stack in P polarization (dotted black), S polarization (dashed black) and the unpolarized case (bold black). For simplicity, on axis incidence is considered only.

The advantage of S polarization over P polarization is evidenced in the two latter examples and it can be justified by comparing the variations of phase $\phi_{RPM}$ -See Appendix A for the definition- in the two polarizations as depicted in Figure 5 for the complete solar cell. It is observed in Figure 5 that phase variations are more considerable in S polarization than in P polarization which is equivalent to higher density of modes and potentially higher absorption.

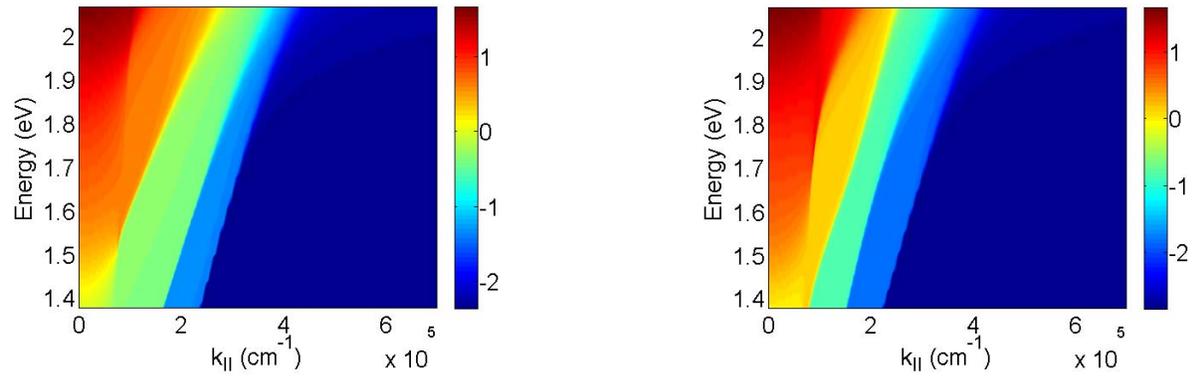

**Figure 5-** (Color online) Variations of $\dfrac{\phi_{RPM}}{\pi}$ versus $k_{\parallel}$ (cm$^{-1}$) and energy (eV) in P (left) and S (right) polarization for a complete solar cell stack. Each mode is associated with a jump of about unit in these graphs.

Two reasons can cause the variation of energy of the guided modes: material absorption and coupling to the other modes due to perturbation of the guided wave structure. Let us suppose that the guided modes are associated to resonance peaks/dips in a corresponding quantity e.g. the absorption or the reflection phase. The quality factors of these resonances can express how much the guided modes are trapped inside the guiding medium. Therefore, sharper resonances correspond to more ideal guided modes which represent low material loss and low rates of energy exchange with the outer media since any sort of energy exchange will necessarily lead to a broadening of the resonance. This argument can explain blurring of the diagrams of Figure 5 at higher energies since the material loss is more considerable in that part. Secondly, since the structure includes thin films and not thick layers, the resonance corresponding to the guided modes will be relatively broad. The latter point can justify blurring of the phase diagrams in Figure 5 at low energies where the layers are weakly absorbing. By going to larger $k_\parallel$, the resonances get sharper since the wave travels mainly parallel to the interfaces; intuitively, the interfaces interact more selectively since the wave passes a longer path from one interface to the other one and its phase is thus more viable to change. Normally, sharp resonances are sensitive to change of their corresponding parameters; that is the reason that in sensing applications they are desired mostly. In the case of solar cells, the cell should not be sensitive to the illumination incidence angle which adds to the benefits of thin film structures.

Up to this point, we have considered 1D gratings only. For the 2D gratings, the same approach as explained can be applied to find the enhancement factor. The mode counting procedure is more complicated but analogous to the 1D grating case and the reciprocal lattice will be two dimensional since the structure will be periodic in two directions. The counting procedure depends on the symmetry of the two dimensional lattice e.g. triangular or square. Figure 6 (a) and (b) show the enhancement factors for the case of a 2D grating with square geometry. The grating period has been assumed 560 nm in both $x$ and $y$ directions. The incident angle is changed by considering two parameters $\theta$ and $\varphi$ which define the incident wave vector as $\vec{k} = |\vec{k}|(\hat{x}\sin\theta\cos\varphi + \hat{y}\sin\theta\sin\varphi + \hat{z}\cos\theta)$. The enhancement factor is plotted versus $\theta$ and $\varphi$; The polar direction defines the $\theta$ axis and the azimuthal direction corresponds to $\varphi$. It is observed that unlike the case of 1D gratings, the maximum potential absorption enhancement using 2D gratings can surpass the $4n^2$ limit regardless of the method of counting, i.e. the "real model" or the "idealized model". Furthermore, the "real model" produces higher enhancement factors than the "idealized model" especially at near normal incidence similar to the case of the 1D grating as a result of smaller effective index. Besides, the angular dependence of the enhancement factor is less if the "idealized" model is applied. Note that according to Figure 6, the enhancement factor shows an anisotropic behavior. It is rather strong along the sides of the Brillouin zone. The Brillouin zones of the square lattice and the triangular lattice geometries are depicted in Figure 6 (c). Figure 6 (d) shows the maximal enhancement factor if a 2D grating with triangular geometry with the lattice constant of 560 nm is used. According to Figure 6 (d), the enhancement factor of cell based on the triangular lattice takes larger values at near normal angles as the case of square geometry and it also shows a rather isotropic behavior as the angle $\varphi$ varies.

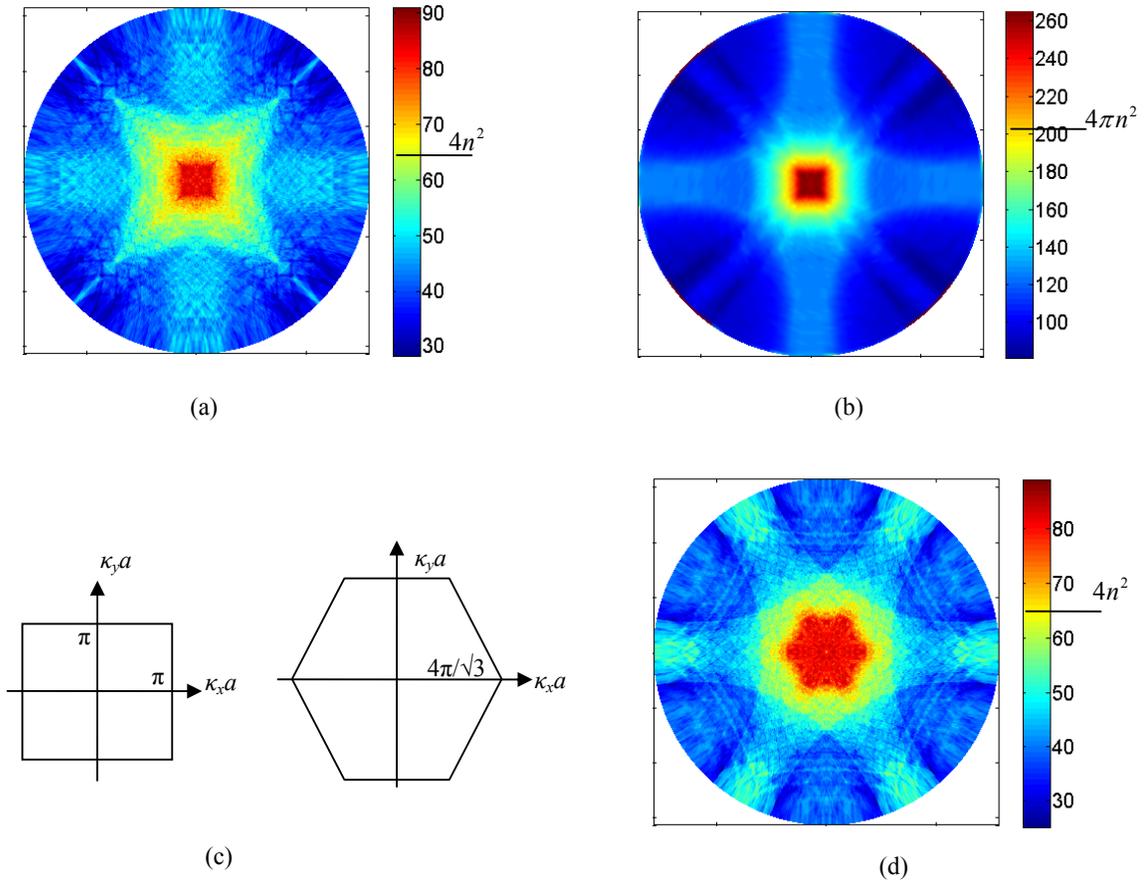

**Figure 6-** (Color online) Top: angular variation of the average enhancement factor under unpolarized light obtained with the "idealized" (a) and the "real" (b) models for the square lattice. The radius represents $\theta$ and the angle shows the corresponding $\varphi$. Bottom: Brillouin zones for the square lattice and triangular lattice geometries (c) and angular variation of enhancement factor obtained with the "idealized" model for the triangular lattice (d).

In summary, thin film structures can provide high enhancement factors if they are optimized effectively by appropriate geometrical design. The approximation of periodicity along the device thickness is not valid for thin films since it does not take into account the open nature of the guided wave structure; it underestimates the enhancement factor. In a real device, there exist other layers than the absorber layer which can cause reduction of the effective refractive index and less enhancement factor. The S polarization provides higher enhancement factor compared to the P polarized one and finally, the 2D periodic thin film devices can potentially surpass the $4n^2$ limit over a wide spectral range.

The enhancement factor definition can sometimes be misleading despite its wide use in the literature; a high value of enhancement factor does not necessarily mean a large photocurrent. The value of $F$ reveals the amount that the absorption can be enhanced by geometrical changes of the device, e.g. texturing it without changing its thickness. If the single path absorption is already high as in thick layers, $F$ will not be very large despite the large amount of absorption. In fact, solar cells with thick layers are expected to absorb more incident energy than their thin film counterparts, however, the high value of F of the thin films can add to their benefits e.g. low material consumption.

## 6. Impact of field overlap: practical limit

The increased effective thickness of the thin film cell is a result of wave spreading outside the structure. But this also means that we can not use the wave energy completely. Even if the wave is totally confined to the cell, the absorption does not occur solely in the active silicon layer; the amount of absorption which occurs outside of silicon should, thus, be considered as parasitic loss since it does not contribute to the photocurrent. The two latter effects, i.e. wave confinement to the cell and parasitic absorption reduce the amount of useful absorption in a solar cell.

The impact of wave's spatial spread and its imperfect confinement to the guide on the enhancement factor was previously discussed in Ref. [16]. In our considerations and generally in the case of high index cells, this effect is marginal [22].

The share of parasitic absorption can be eliminated from the total absorption by defining an "absorption overlap" $\eta_{abs}$ as the ratio of the absorption in the active layer of the cell to the absorption integral of the whole cell. One finds that

$$\eta_{abs} = \frac{\int_{i-si} \varepsilon_i |\mathbf{E}|^2 dV}{\int_{cell} \varepsilon_i |\mathbf{E}|^2 dV} . \tag{4}$$

where $dV$ indicates the volume element. The overall absorption enhancement factor can be obtained by multiplying F with the latter overlap integral.

$$F_{eff} = \eta_{abs} F . \tag{5}$$

The modification of F in Eq. (5) is general and not restricted to a special geometry. To evaluate it, it is necessary to have the electric field profile of the mode under different coupling conditions, i.e. $(k_{\parallel}, E)$ where $E$ is the photon energy. Since the structures investigated in this manuscript are relatively flat, the electric field profiles can be in good faith approximated with the ones of a flat interface structure. Furthermore, for each resonance, the mode profile at the resonance peak is used.

Reduction of enhancement factor as a result of the energy overlap with the device geometry and the absorption distribution is not found only in thin films. It is not even constrained to guided wave geometries. In a more general view, whenever there is confinement of energy or resonance, there will be leakage and loss of that energy. The amount of energy leakage and loss determines the quality factor of the resonance. High quality factor resonances provide highly confined waves and sharp spectral features. For example, in cells with thick layers, the resonances are rather sharp and the wave is almost confined in the structure. However, it should be noticed that highly confined modes do not necessarily support a high amount of absorption enhancement in the solar cell. An example of the latter case is a plasmonic mode which is localized at the metal-dielectric interface but a huge portion of its corresponding absorption occurs in the metal; therefore, its $\eta_{abs}$ and $F_{eff}$ are small. Thus, absorption in the metal is a major challenge for applying plasmonics to thin film solar cells [23]. Figure 7 shows the effect of the absorption overlap on the maximal enhancement factor of the complete thin film solar cell stack described in the section 5 where a 1D grating coupler with P = 560 *nm* is used. By applying the "absorption overlap", the enhancement drops drastically to almost its half. Therefore, the parasitic loss is a major factor which limits the enhancement factor.

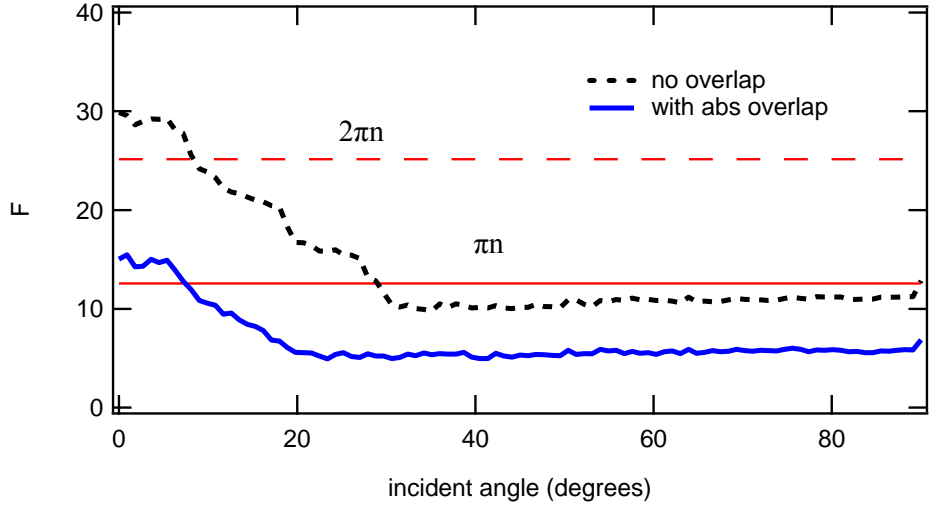

**Figure 7-** (Color online) Effect of loss on the average enhancement factor of the full solar cell stack with 1D grating coupler (P=560nm) under unpolarized illumination. The curve with "no overlap" is the benchmark where modification of relation (5) is not applied. The "with abs overlap" corresponds to the case where this modification is applied. For simplicity, on axis incidence is considered only.

## *7. Practical considerations*

So far we have obtained the average absorption enhancement factor over the wavelength range from 600 to 900 nm. In this section we calculate the maximal increase in the short circuit current density ($J_{sc}$) in the same range provided by guided modes. To do so, we obtain primarily the absorption via $A = F\alpha d$ and subsequently the increase in $J_{sc}$ by a weighted integration of absorption over wavelength. In this way we provide a better approximation of the absorption enhancement than the simple averaging of $F$ over wavelength since we take into account also the solar flux density.

Figure 8 shows the increase in $J_{sc}$ under unpolarized illumination versus the incident angle for the complete solar cell stack of sections 5 and 6 on a 1D grating coupler with P=560*nm*. The dotted curve shows the $J_{sc}$ increase assuming that the wave is fully confined to the i-Si layer ($\eta_{abs} = 1$). The solid curve considers the absorption overlap defined in Eq. (5) i.e. it takes into account the impact of parasitic absorption. The maximal photocurrent increase is obtained still at near normal incidence. If the absorption occurs only in the i-Si layer, the guided modes can increase the photocurrent at most by 2 mA/cm$^2$. However, absorption in the other layers limits this photocurrent increase to 1 mA/cm$^2$. This confirms again the undesirable effect of parasitic absorption. Even if one uses 2D gratings or random textures to introduce more mode excitations and thus more photocurrent enhancement, parasitic absorption will be a bottleneck.

It is worth mentioning that for a cell encapsulated beneath a planar glass sheet, the angle of incidence on the glass-ITO interface will be in a narrow range near normal. The glass layer also polarizes the incident light. It even changes the whole dispersion diagram of the multilayer stack. However, the angular behavior of $F$ and $J_{sc}$ should resemble the case without glass at least qualitatively. The reason is that F depends not only on the incident angle but also on the refractive index of the incident medium. Limiting the incidence to small angles by increasing the index of the incidence medium does not increase F due to radiance theorem. To study more complicated effects e.g. polarization, one should use a modified dispersion diagram which considers also the glass layer. The unpolarized result should be anyway obtained by simple averaging of the P- and S-polarized results in this case as before.

Although the absorption enhancement depends tightly on the properties of a cell's layers, we can still provide some guidelines for the design of interface textures. Generally, 2D gratings or random textures are preferred to 1D gratings since they support more guided mode excitations. Specifically 2D gratings with triangular geometry can also reduce the dependence of the absorption on the incident angle.

Apart from the mentioned points, the method which we used to obtain the dispersion diagram and to find spread of the modes over $(k_\parallel, E)$ plane can be used to analyze guided modes in thin film solar cells. Therefore, by using the mentioned technique, it is possible to engineer the guided modes for a cell with an arbitrary number of layers and materials. This can be considered as an initial but important step in the design of textured solar cells.

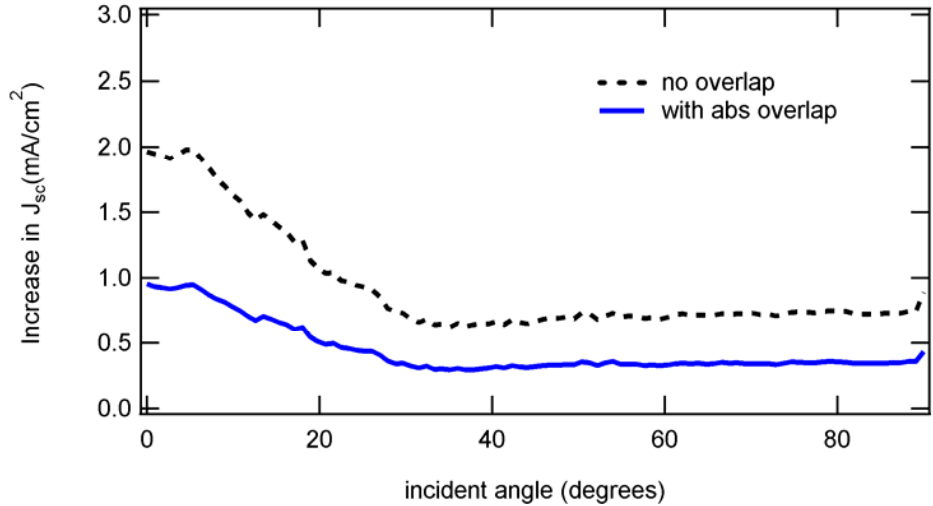

**Figure 8- (Color online) Angular dependence of photocurrent increase for the full solar cell stack with a 1D grating coupler with P=560 nm under unpolarized light. The curve with "no overlap" is where modification of Eq. (5) is not applied. The "with abs overlap" corresponds to the case where this modification is applied. For simplicity, on axis incidence is considered only.**

## *8. Conclusion*

The angular behavior of the upper limit of light trapping in thin film solar cells was investigated theoretically in this work. In this contribution, we showed that by benefiting from guided modes, it can be possible to enhance the effective light path length in thin film solar cells beyond the wide band $4n^2$ limit over a wide angular range and beyond the $4\pi n^2$ limit in a relatively narrow angular range, both over a wide wavelength range. We discussed the different types of optical resonances in the solar cell and their angular behavior. Absorption enhancement was calculated in this contribution by two models; in the "idealized model", field penetration into air was neglected but in the "real model", it was taken into account by considering the dispersion diagram of the planar structure. To be able to count the modes we established a procedure in which we found the modes by looking at the reflection phase. We considered different cases in our examples: a slab waveguide made of a material with n=4, the same geometry with refractive index dispersion of amorphous silicon and a complete solar cell stack. We showed that the absorption in an exemplary thin film solar cell can be increased by a factor of 30 if a 1D grating texture is used. This absorption enhancement factor can go up to 260 for a 2D grating with square geometry. We showed that with a triangular 2D lattice, it is possible to reduce the angular dependence of the absorption enhancement. Our calculations include the effect of wave spread outside the solar cell on the enhancement factor. Furthermore, we considered a further reduction in the enhancement factor introduced by the parasitic loss in the cell layers; this helped us to obtain a more practical absorption enhancement limit. By inclusion of that effect, the enhancement factor dropped to almost half of its initial value. Although there is a trade-off between the amount of energy confinement and the number of modes, the major limiting mechanism for the absorption enhancement was found to be the parasitic loss. Finally, we calculated the maximal photocurrent increase of 2 mA/cm$^2$ provided by a 1D grating coupler for a full solar cell stack. If we consider also the effect of parasitic absorption, this photocurrent enhancement is reduced to 1 mA/cm$^2$.

# Appendix A: Counting the resonances of the flat interface structure

The Transfer Matrix Method (TMM) and the Reflection Pole Method (RPM) [24] are used to calculate the reflection and the dispersion diagrams respectively. Using TMM, forward and backward fields in the incident medium (air) and the transmission medium (silver) can be related with a simple 2 by 2 matrix Q.

$$\begin{pmatrix} F_t \\ B_t \end{pmatrix} = Q \begin{pmatrix} F_i \\ B_i \end{pmatrix} = \begin{pmatrix} q_{11} & q_{12} \\ q_{21} & q_{22} \end{pmatrix} \begin{pmatrix} F_i \\ B_i \end{pmatrix}. \quad \text{(A-1)}$$

In the above relation, $F$ and $B$ stand for the forward and backward waves respectively and the indices $i$ and $t$ correspond to the incident or the transmission medium respectively. Without losing generality we assume all the waves normalized to the incident wave, i.e. $F_i = 1$. Since light does not illuminate the structure from the bottom, $B_t = 0$. Therefore, Eq. (A-1) reduces to the system of two equations and two unknowns which can be solved very easily. Specifically, reflection from the top interface can be represented as

$$r_c = |r_c| \exp(j\phi_c) = -\frac{q_{21}}{q_{22}}. \quad \text{(A-2)}$$

Theoretically, poles of the reflection coefficient $r_c$ correspond to the guided modes. As evident in Eq. (A-2), reflection becomes infinite when $q_{22} = 0$. The RPM monitors the phase $\phi_{RPM}$ of $q_{22} = |q_{22}| \exp(j\phi_{RPM})$ when the longitudinal wave vector, $k_\parallel$, is changing. In the case of pure dielectric waveguides, $\phi_{RPM}$ will change in steps equal to $\pi$ exactly at the $k_\parallel$ values corresponding to propagation constants of the waveguide modes. If loss is introduced into the structure as well, $\phi_{RPM}$ varies abruptly by about $\pi$ around the propagation constants of the modes [24]. So, an "effective" number of modes can be found for each polarizations by the dividing amount of variation of $\phi_{RPM}$ by $\pi$. The word "effective" is used since in thin films, the individual resonances are relatively wide compared to the case of thick layers. Therefore, it is not always possible to distinguish different resonances since at each wavelength, there may be more than one resonance which enhances absorption.

To find the number of allowed resonances, $\Delta M$, in each energy range $[E, \Delta E]$, we should consider limitations imposed by both the $\phi_{RPM}$ and the periodicity $P$ along $x$ direction. The latter criterion necessitates that

$$k_\parallel = k_0 \sin\theta + \frac{2\pi m}{P}. \quad \text{(A-3)}$$

which corresponds to vertical lines in the $k_\parallel - E$ plane of Figure 1. Note that we take into account only parts of the phase diagrams which correspond to $k_\parallel > k_0$ to consider only the guided modes.

# Appendix B: Calculation of the enhancement factor in the "real model"

The process of finding the absorption enhancement factor, F, via the "real model" can be described as follows:
1) Divide the energy range into tiny subintervals $\delta E$.

2) Move on the curve of +1 diffraction order, i.e. $m=1$ in Eq. (A-3), and find the change of $\phi_{RPM}$ (introduced in Appendix A) on this curve in each energy subinterval $\delta E$. Divide this phase shift by $\pi$ to obtain the effective number of modes $\delta m$ in $\delta E$.

3) Divide $\delta m$ by the average phase index ($n_p$) on the curve $m=1$ over $\delta E$.

4) Repeat the previous steps for different diffraction orders which excite the guided modes in the spectral range of interest.

4) Add $\delta E/n_p$ for different diffraction orders at each $\delta E$ to obtain a total $\delta E/n_p$ which depends only on the photon energy and not on the diffraction order.

5) Find the number of diffraction orders, $N$, at each $\delta E$.

6) Calculate the enhancement factor $F$ via $F = \dfrac{c}{Nd}\dfrac{\delta m}{n_p}$ and average it over the whole wavelength range of interest to find the average enhancement factor.

## Acknowledgement

We thankfully acknowledge the funding by the Swiss National Science Foundation under project number 200020_137700/1.